\documentclass[conference]{IEEEtran}
\ifCLASSINFOpdf
\else
\fi

\ifCLASSOPTIONcompsoc
    \usepackage[caption=false, font=normalsize, labelfont=sf, textfont=sf]{subfig}
\else
	\usepackage[caption=false, font=footnotesize]{subfig}
\fi

\usepackage{cite}
\usepackage[version=4]{mhchem}
\usepackage{graphicx}
\usepackage{amsmath,amssymb,amsfonts}
\usepackage{algorithm}
\usepackage[noend]{algpseudocode}
\usepackage{tablefootnote}
\usepackage[table,xcdraw]{xcolor}
\usepackage{lipsum}
\usepackage{textcomp}
\usepackage{tikz}
\usepackage[utf8]{inputenc}
\usepackage{pgfplots} 
\usepackage{pgfgantt}
\usepackage{pdflscape}
\pgfplotsset{compat=newest} 
\pgfplotsset{plot coordinates/math parser=false}
\usetikzlibrary{plotmarks}

\begin{document}
\title{CrossStack: A 3-D Reconfigurable RRAM Crossbar Inference Engine}

\author{Jason~K.~Eshraghian$^1$,~\IEEEmembership{Member,~IEEE,}
		Kyoungrok~Cho$^2$~\IEEEmembership{Senior~Member,~IEEE,}
        and~Sung~Mo~Kang$^3$~\IEEEmembership{Life~Fellow,~IEEE,}\\%
        
        \IEEEauthorblockA{$^1$\textit{School of Electrical, Electronic and Computer Engineering, University of Michigan, Ann Arbor, MI 48109 USA}}
        \IEEEauthorblockA{$^2$\textit{College of Electrical and Computer Engineering, Chungbuk National University, Cheongju 362763, South Korea}}
\IEEEauthorblockA{$^3$\textit{Jack Baskin School of Engineering, University of California, Santa Cruz, Santa Cruz, CA 95064 USA}}}%


\maketitle

\begin{abstract}
Deep neural network inference accelerators are rapidly growing in importance as we turn to massively parallelized processing beyond GPUs and ASICs. The dominant operation in feedforward inference is the multiply-and-accumlate process, where each column in a crossbar generates the current response of a single neuron. As a result, memristor crossbar arrays parallelize inference and image processing tasks very efficiently. In this brief, we present a 3-D active memristor crossbar array `CrossStack', which adopts stacked pairs of \ce{Al/TiO2}/\ce{TiO_{2-x}}/\ce{Al} devices with common middle electrodes. By designing CMOS-memristor hybrid cells used in the layout of the array, CrossStack can operate in one of two user-configurable modes as a reconfigurable inference engine: 1) expansion mode and 2) deep-net mode. In expansion mode, the resolution of the network is doubled by increasing the number of inputs for a given chip area, reducing IR drop by 22\%. In deep-net mode, inference speed per-10-bit convolution is improved by 29\% by simultaneously using one \ce{TiO2}/\ce{TiO_{2-x}} layer for read processes, and the other for write processes. We experimentally verify both modes on our $10\times10\times2$ array.
\end{abstract}



\section{Introduction}
\IEEEPARstart{I}{ncreasing} the sizes of artificial neural networks (ANNs) has been the most common response to the copious amount of data being continuously generated. Where training sets are in excess of billions of inputs processed through hundreds of millions of parameters in a neural network \cite{Mahajan2018}, new ways to speed up the processing of all this information must be developed. Since 2012, the training runtime of neural networks has doubled every 3--4 months. It is equally important to develop hardware that is not only optimized for running very large scale networks, but is adaptable to the unceasing wave of emerging ANN topologies.

Memristors are now ubiquitous in neuromorphic computing literature due to their long retention \cite{Kim2010, Kumar2017, Lin20201}, excellent scalability \cite{Pi2019, Fuller2019}, fast read and write speeds \cite{Merced2016, Lin20202}, compatibility with CMOS technology \cite{Eshraghian2018, Chakrabarti2017, Cai2019, Azghadi2020}, and precise weight updates \cite{Lammie2020, Krestinskaya2018}. The development of dense integrated structures with 3-D stacked crossbar arrays enables an increase in the throughput for a given chip area, but thus far, target applications of 3-D RRAM have mostly been limited to digital memory \cite{Chevallier2010, Wang2011, Jo2014, Hong2018, Azghadi20201, Baek2019, Fastow2019, Eshraghian2017}.

It can be difficult for ASIC designs to keep pace with the latest developments in machine learning due to the lag time between algorithm development and the full IC design cycle. With popular machine learning methods in a rapidly evolving state, reconfigurability is of paramount importance to ensure hardware does not become obsolete the moment new network architectures and topologies are introduced. In response to this, we present a reconfigurable stacked pair of memristor crossbars dubbed `CrossStack' that can be operated in one of two modes: 1) expansion mode, and 2) deep-net mode. In expansion mode, the resolution of the network is doubled by increasing the number of inputs for a given chip area, thus reducing IR drop by 22\% of an equivalent array. In deep-net mode, inference speed per-10-bit convolution is improved by 29\% by simultaneously using one array for read processes, and the other for write processes. This brief will demonstrate how to selectively isolate and couple the two layers using CMOS cell design. We experimentally verify this on our in-house fabricated crossbar stack, using separately controlled CMOS circuitry in the SK Hynix 180nm process. 

\section{Matrix-Vector Multiplication}
To perform analog-domain multiply-and-accumulate (MAC) using RRAM arrays, 
a voltage vector $\boldsymbol V^i$ is applied at the input, multiplied by a conductance matrix $\boldsymbol G$, to generate a current vector of $\boldsymbol i=\boldsymbol V^i \boldsymbol G$ in accordance with Ohm's Law and Kirchhoff's Current Law. On a pre-trained network, the conductance of each memristor is programmed prior to read-out. For further detail on MAC on a crossbar, we recommend referring to \cite{Hu2016,Shafiee2016}. In a conventional crossbar, the memristors must be programmed prior to read-out. Where the number of parameters exceed the memory resources available, RRAM cells must be reprogrammed while the data flow of the activations is stalled. 
CrossStack avoids possible stalling that may occur by adding the option to pipeline the read and write processes simultaneously across the two layers. How this is achieved will be described in the following section.

\section{Circuit Operation Modes}

\begin{figure}
\centering
\subfloat[]
{
	\includegraphics[scale=.7]{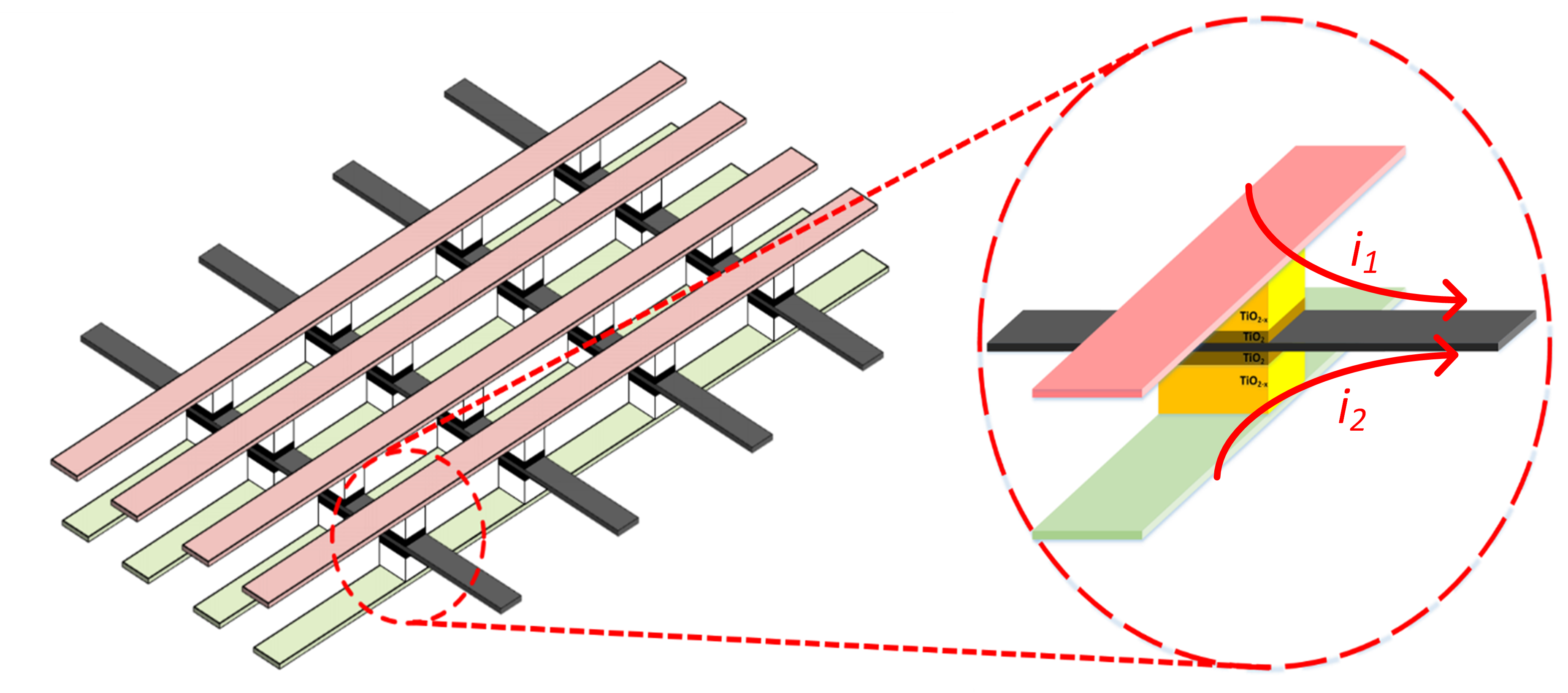}
	\label{fig:2a}
}
\\
\subfloat[]
{
	\includegraphics[scale=.38]{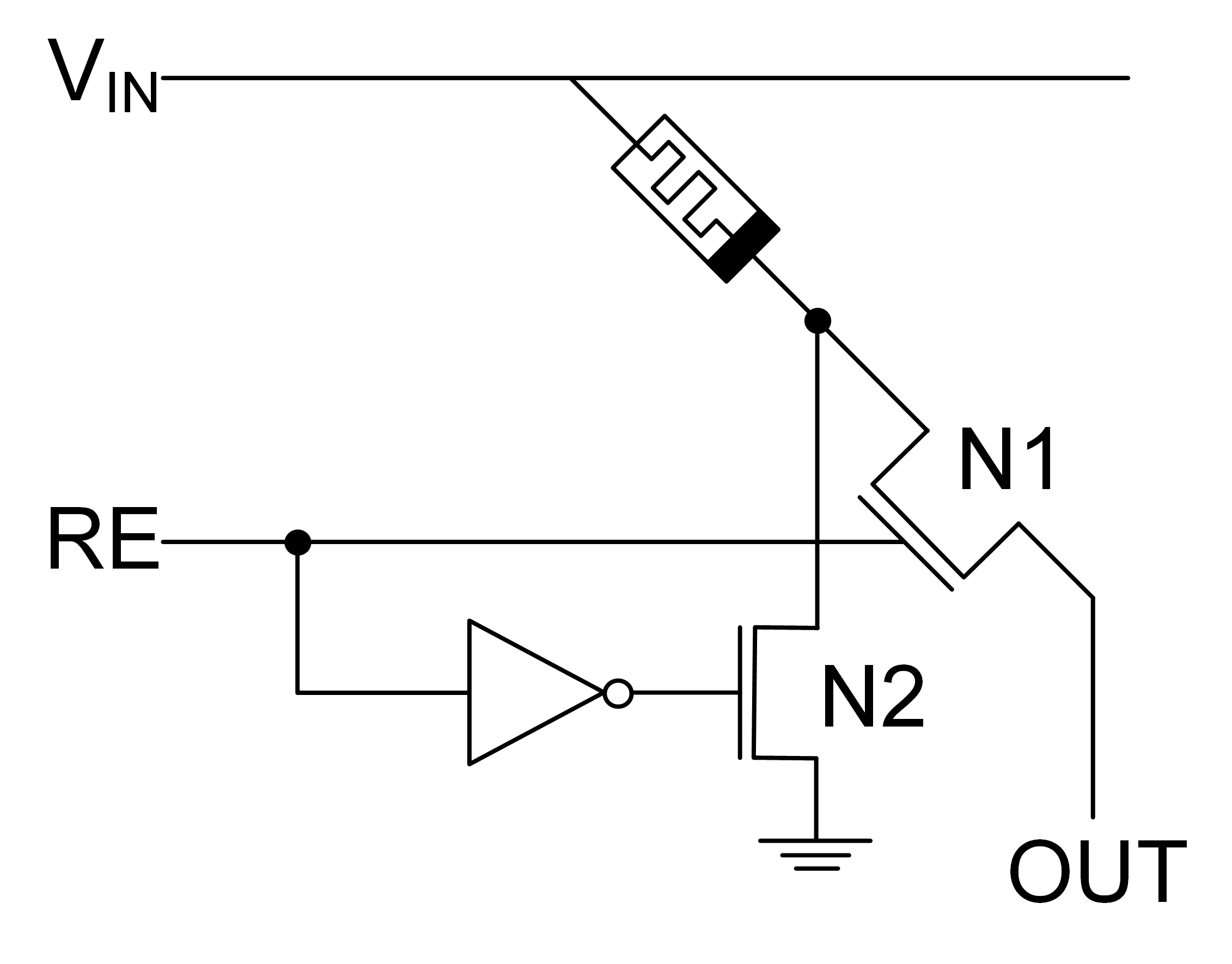}
	\label{fig:2b}
}
\subfloat[]
{
	\includegraphics[scale=.38]{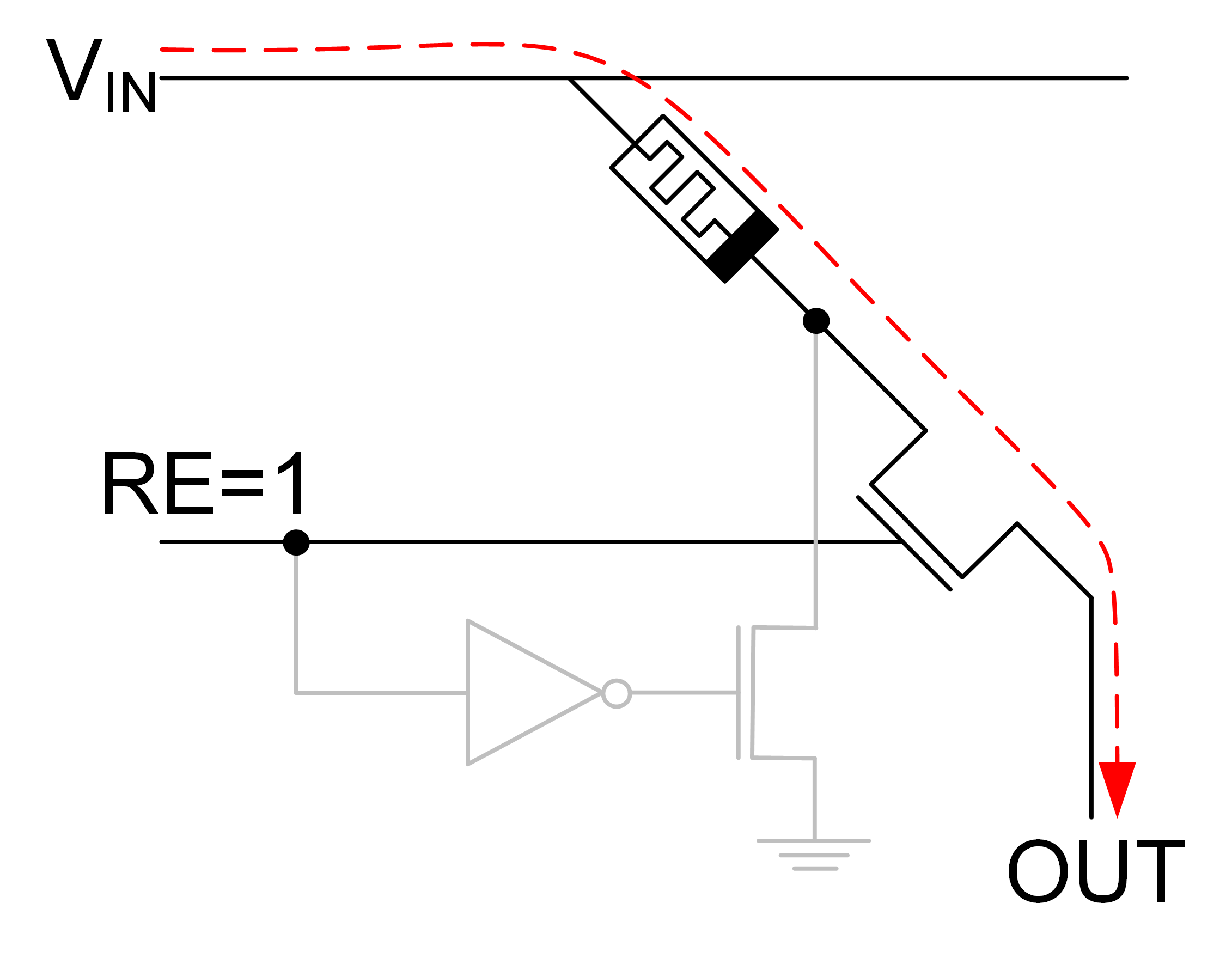}
	\label{fig:2c}
}
\subfloat[]
{
	\includegraphics[scale=.38]{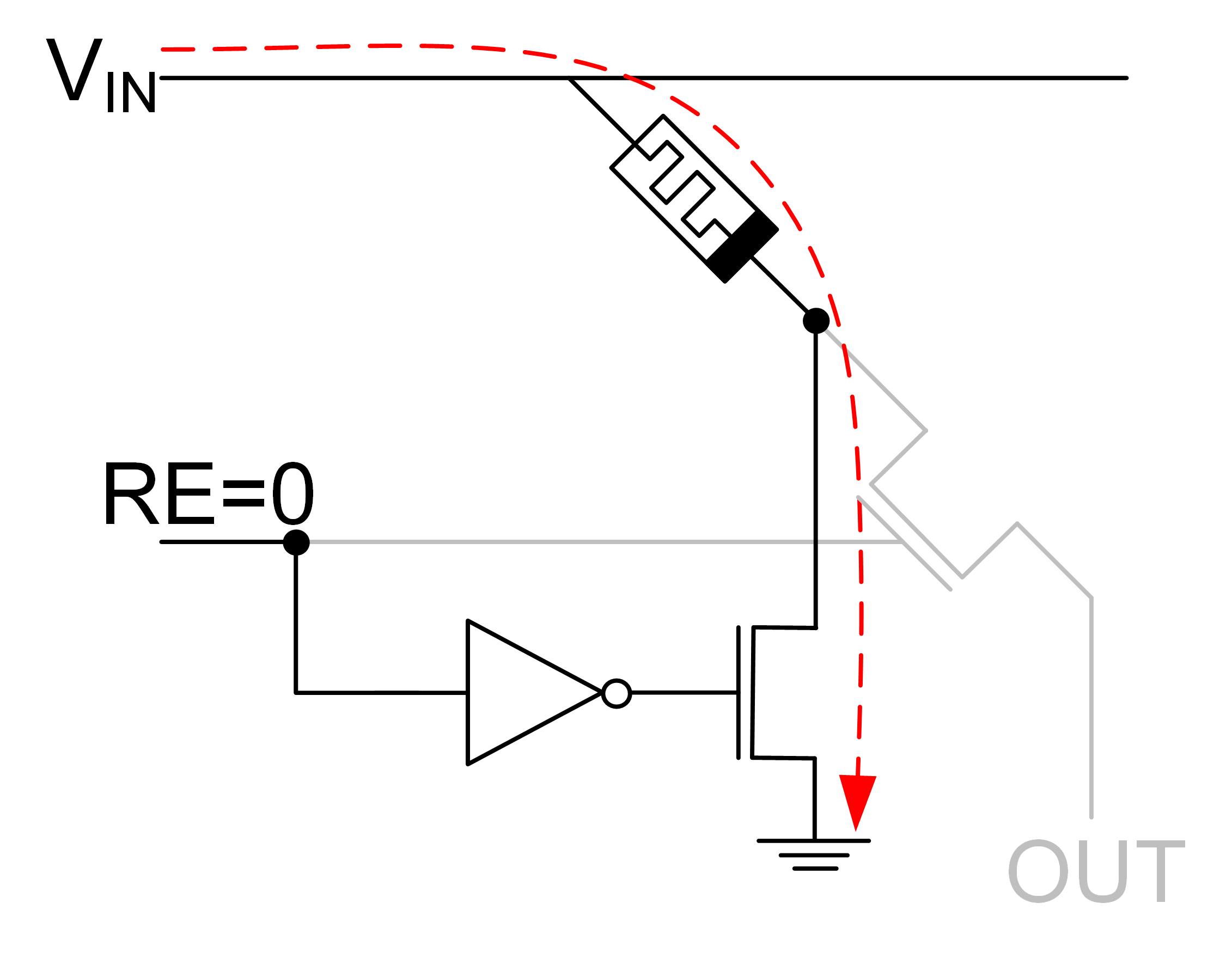}
	\label{fig:2d}
}\\
\subfloat[]
{
	\includegraphics[scale=.55]{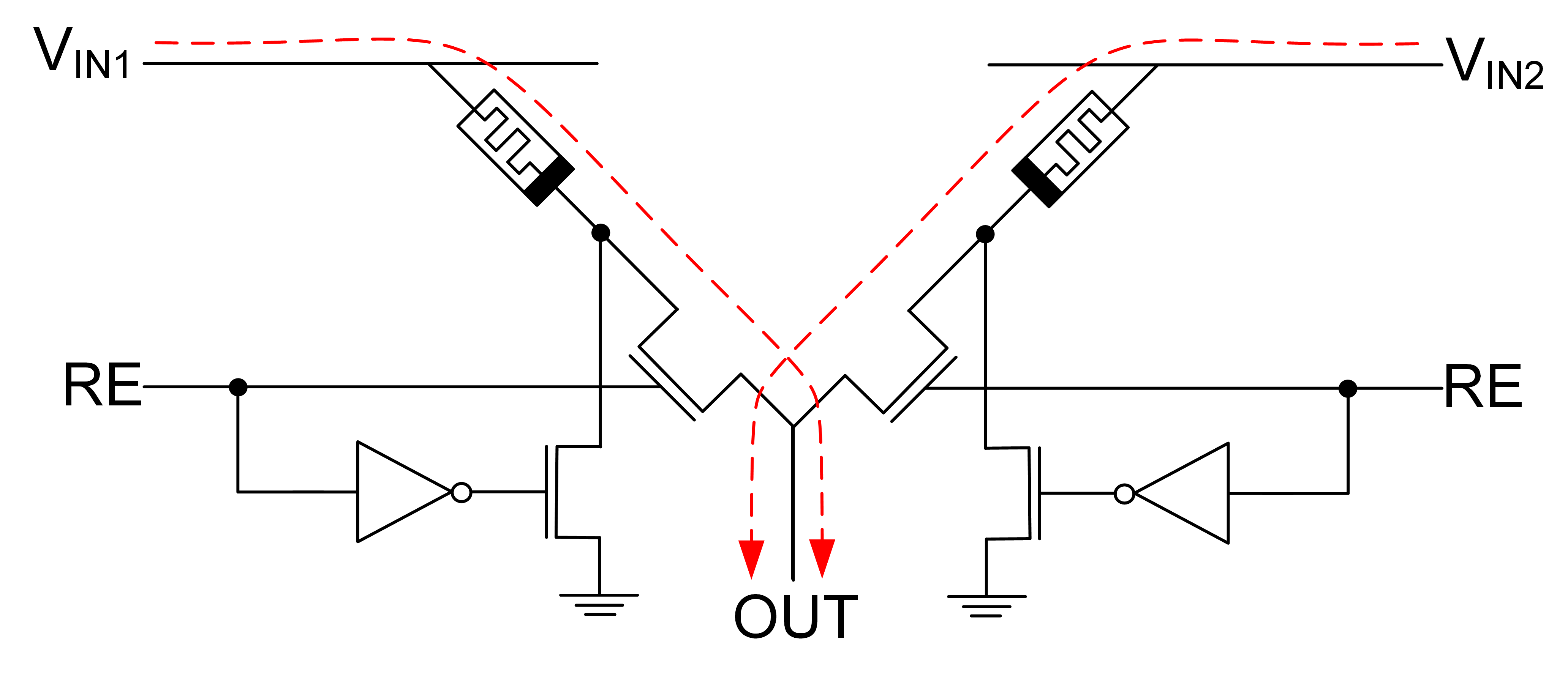}
	\label{fig:2e}
}\\
\subfloat[]
{
	\includegraphics[scale=.55]{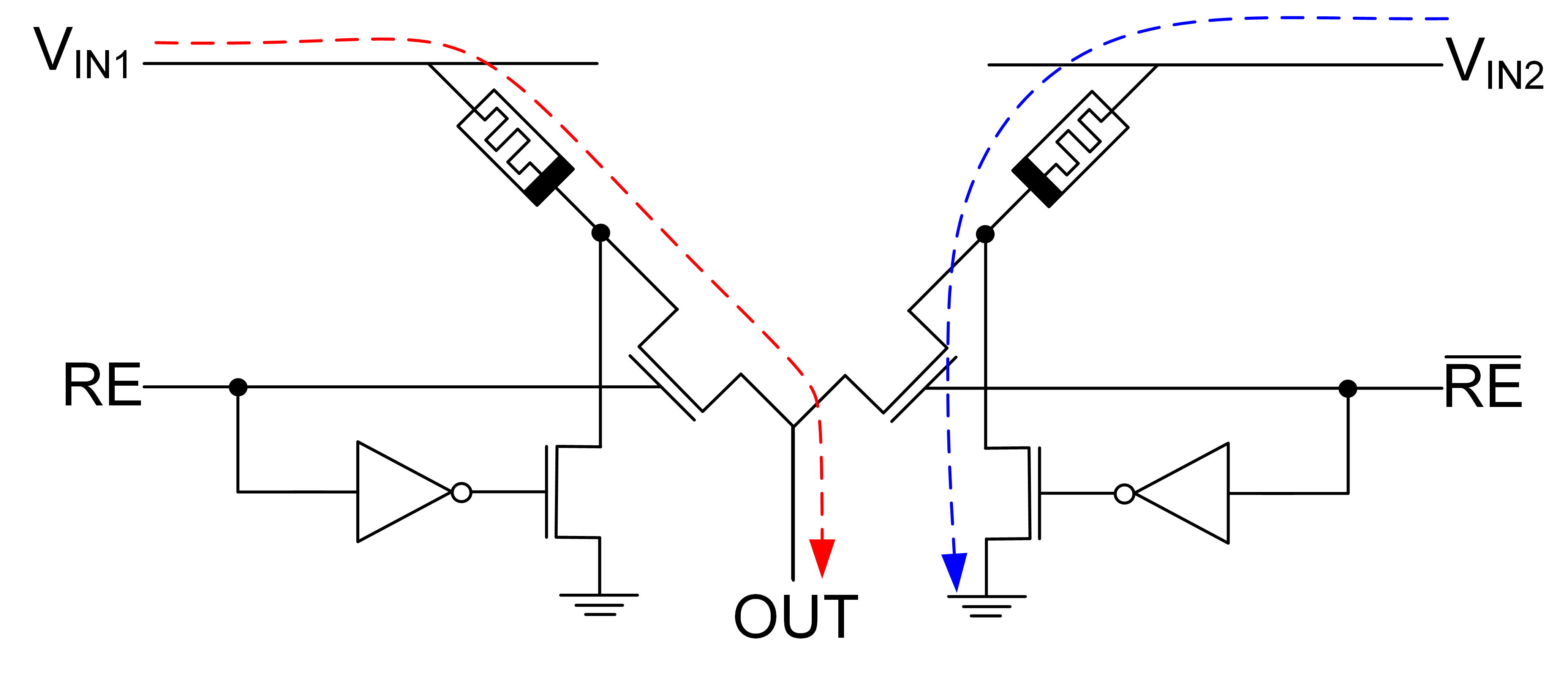}
	\label{fig:2f}
}
\caption{(a) Simplified 3-D memristor crossbar array in expansion mode with cumulative current through the shared electrode (b) Memristor-CMOS cell (c) Current flow in read mode when read-enable is set high (d) Current flow in write mode when read-enable is set low (e) A stacked pair of cells during the read cycle in expansion mode (f) A stacked pair of cells in deep-net mode.}
\label{fig:2}
\end{figure}

A simplified structure of CrossStack is depicted in Fig.~1(a), and the memristor-CMOS cell schematic is given in Fig.~1(b). This work presents two modes in which CrossStack may operate in. These modes are controlled by an active-high read-enable signal \texttt{RE}.

\subsection{Expansion Mode} 
Expansion mode enables access to a shared column line from memristors both above and below the wire. Vertical stacking of memristors doubles the number of possible inputs and weights to each neuron for a given length of column wire which is illustrated in Fig. 1(a) when compared to conventional crossbars. This can be formalized by the following equation:

\begin{equation}
\label{Eq1}
\begin{bmatrix} 
i_1 & i_2 & ... & i_m
\end{bmatrix} = 
\begin{bmatrix} 
V_1^i \\ V_2^i \\ \vdots \\ V_n^i 
\end{bmatrix}^T 
\begin{bmatrix} 
G_{1,1} & G_{1,2} & ... & G_{1,m} \\
G_{2,1} & G_{2,2} & ... & G_{2,m} \\ 
\vdots & \vdots & \ddots & \vdots \\ 
G_{n,1} & G_{n,2} & ... & G_{n,m}
\end{bmatrix}
\end{equation}

\noindent where $m$ is the number of columns in the crossbar and $n$ is the number of rows. In expansion mode, $n$ is double that of a 2-D array, as there are rows both above and below the shared column contributing to output current.



To activate a cell in expansion mode, the read-enable signal of all cells must be identical. The current pathway from input to output of a single cell is depicted in Fig.~1(c), and a pair of stacked cells is shown in Fig.~1(e). To generate a read-out current at each shared column, \texttt{RE} must be set high (in our case, $V \geq V_{Th} = 0.4 V$; described in further detail in our experimental results). If transistors \texttt{N1} and \texttt{N2} are treated as switches, then \texttt{N2} would be off and \texttt{N1} would be on. Therefore, a current pathway is formed from both upper and lower crossbar arrays to the column line. To program a memristor, \texttt{RE} is set low as in Fig.~1(d). This causes \texttt{N2} to switch on and \texttt{N1} off, which forms a pathway from the memristor to ground and prevents current from flowing to the shared column. Therefore, the two crossbars can be programmed independently of one another by isolating them. The transistors are sized to ensure a negligible leakage current, and to sustain a sufficiently low \texttt{ON} resistance in comparison to the memristance while it is operating in the linear region such that it behaves as an ideal switch.

\subsection{Deep-net Mode}


Deep-net mode ensures both pairs of arrays are isolated from one another at all times. Isolation biasing enables each layer to operate independently. The two arrays must have complementary \texttt{RE} signals as distinct from expansion mode, depicted in Fig.~1(f). This means that while one array generates a read-out current, the other array of memristors are being programmed (in write mode). As described above, when \texttt{RE} is low the cell does not contribute to read-out current and input voltage $V^{i}$ for the write layer is applied such that the conductances written to the memristors correspond to the weights of the next hidden layer in the neural network. Once the analog output current is digitized as a voltage, the write-layer has been pre-programmed and there is no need to buffer the current or store it in memory, as is required by most other pipelines \cite{Shafiee2016, Song2017, Valavi2019}. This process is repeated, but now the roles of the crossbars are reversed. The original read-layer is now programmed to the next hidden layer of weights, and the original write-layer generates the read-out current. Thus, read-write processes run in parallel. One layer is programmed in anticipation of the output from the other layer, which enables a novel in-situ pipeline.

In describing how to program a memristor, the key difference between expansion and deep-net modes is that in expansion, all cells are identically biased for either read or write at any given time. In deep-net mode, 50\% of the cells are biased for read processes and the remaining 50\% are biased for write processes, only switching once each operation is complete. The shorter read-out time is subsumed within the programming cycle, but at the expense of half the number of inputs $n$ in (1). We quantitatively demonstrate this in our experimental results.

\section{Experimental Results}

\subsection{Crossbar Fabrication}

\begin{figure}
\includegraphics[scale=1]{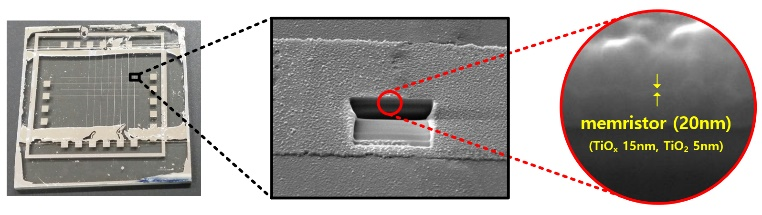}
\caption{A $10 \times 10 \times 2$ prototype of CrossStack, with a cross-sectional view of the active layer taken using a focus ion beam analyzer.}
\end{figure}
CrossStack was constructed based on two monolithically integrated crossbar arrays with a shared central electrode which make up the column line, based on a sandwich structure of \ce{Al}/\ce{TiO2}/\ce{TiO_{2-x}}/\ce{Al} layers. A layer of Al (200-nm-thick and 20-$\mu$m-wide) was deposited using photolithography on a glass wafer as the bottom electrode (irradiated using mask alignment for 100~s, subsequently developed at  23$^{\circ}$C for 120~s). Any excess Al outside of the channel region was removed via wet etching (\ce{H3PO4} : \ce{HNO3} : \ce{CH3COOH} : \ce{H2O} = 80 ml : 5 ml : 5 ml : 10 ml) at a rate of $\Delta$d/t = 300 nm/min. \ce{TiO2} (5-nm-thick) and \ce{TiO_{2-x}} (15-nm-thick) thin films were formed by atomic layer deposition and magnetron sputtering to fabricate the memristor. Another 200-nm-thick layer of Al was sputtered as the top electrode using photolithography to create a 20~$\mu$m $\times$ 20~$\mu$m mask. After a planarization step, the top stack of active layer and metal were also deposited. Note that the polarity of the pair of active layers are mirrored, as distinguishable from \cite{Adam2017}. This allows for identical input voltages to be applied when programming the memristors. Fig.~2 shows a working $10 \times 10 \times 2$ prototype of CrossStack, and a cross-sectional view of the memristor taken using a focus ion beam analyzer.

\begin{table}[]\scriptsize
\centering\caption{CrossStack Characteristics}
\begin{tabular}{
>{\columncolor[HTML]{FFFFFF}}l 
>{\columncolor[HTML]{FFFFFF}}l 
>{\columncolor[HTML]{FFFFFF}}l }
\hline\vspace{-6pt}
 &  &  \\ \hline
\multicolumn{1}{|c|}{\cellcolor[HTML]{FFFFFF}Symbol} & \multicolumn{1}{l}{\cellcolor[HTML]{FFFFFF}Parameter} & \multicolumn{1}{r|}{\cellcolor[HTML]{FFFFFF}Value} \\ \hline
\multicolumn{1}{|c|}{\cellcolor[HTML]{FFFFFF}$R_s$} & \multicolumn{1}{l}{\cellcolor[HTML]{FFFFFF}static resistance of set; \color{red}\textit{$R_s=\frac{1}{g_s}$}} & \multicolumn{1}{r|}{\cellcolor[HTML]{FFFFFF}10K$\Omega~\pm$ 7\%} \\
\multicolumn{1}{|c|}{\cellcolor[HTML]{FFFFFF}$R_{r}$} & \multicolumn{1}{l}{\cellcolor[HTML]{FFFFFF}static resistance of reset;\color{red}\textit{$R_r=\frac{1}{g_r}$}} & \multicolumn{1}{r|}{\cellcolor[HTML]{FFFFFF}100K$\Omega~\pm$ 10\%} \\
\multicolumn{1}{|c|}{\cellcolor[HTML]{FFFFFF}$V_{DD}$} & \multicolumn{1}{l}{\cellcolor[HTML]{FFFFFF}supply voltage} & \multicolumn{1}{r|}{\cellcolor[HTML]{FFFFFF}1.8 V} \\
\multicolumn{1}{|c|}{\cellcolor[HTML]{FFFFFF}$V_{read}$} & \multicolumn{1}{l}{\cellcolor[HTML]{FFFFFF}read voltage} & \multicolumn{1}{r|}{\cellcolor[HTML]{FFFFFF}0.5 V} \\
\multicolumn{1}{|c|}{\cellcolor[HTML]{FFFFFF}$V_{write}$} & \multicolumn{1}{l}{\cellcolor[HTML]{FFFFFF}write voltage} & \multicolumn{1}{r|}{\cellcolor[HTML]{FFFFFF}1.2 V} \\
\multicolumn{1}{|c|}{\cellcolor[HTML]{FFFFFF}$t_{read}$} & \multicolumn{1}{l}{\cellcolor[HTML]{FFFFFF}current-read out time} & \multicolumn{1}{r|}{\cellcolor[HTML]{FFFFFF}10 ns} \\
\multicolumn{1}{|c|}{\cellcolor[HTML]{FFFFFF}$t_{write}$} & \multicolumn{1}{l}{\cellcolor[HTML]{FFFFFF}programming time} & \multicolumn{1}{r|}{\cellcolor[HTML]{FFFFFF}250 ns} \\ 
\multicolumn{1}{|c|}{\cellcolor[HTML]{FFFFFF}$n$} & \multicolumn{1}{l}{\cellcolor[HTML]{FFFFFF}number of memristors} & \multicolumn{1}{r|}{\cellcolor[HTML]{FFFFFF}200} \\
\multicolumn{1}{|c|}{\cellcolor[HTML]{FFFFFF}$v_{Th}$} & \multicolumn{1}{l}{\cellcolor[HTML]{FFFFFF}threshold voltage} & \multicolumn{1}{r|}{\cellcolor[HTML]{FFFFFF}$|$0.4$|$ V} \\ 
\multicolumn{1}{|c|}{\cellcolor[HTML]{FFFFFF}$P_{critical}$} & \multicolumn{1}{l}{\cellcolor[HTML]{FFFFFF}worst case power consumption} & \multicolumn{1}{r|}{\cellcolor[HTML]{FFFFFF}2.9 mW} \\ 
\multicolumn{1}{|c|}{\cellcolor[HTML]{FFFFFF}$R_{wire}$} & \multicolumn{1}{l}{\cellcolor[HTML]{FFFFFF}wire resistance} & \multicolumn{1}{r|}{\cellcolor[HTML]{FFFFFF}3.2$\Omega$ p/cell} \\ 
\multicolumn{1}{|c|}{\cellcolor[HTML]{FFFFFF}$A_{cell}$} & \multicolumn{1}{l}{\cellcolor[HTML]{FFFFFF}cell area} & \multicolumn{1}{r|}{\cellcolor[HTML]{FFFFFF}20 $\mu$m $\times$ 20$\mu$m} \\ 
\multicolumn{1}{|c|}{\cellcolor[HTML]{FFFFFF}$W/L$} & \multicolumn{1}{l}{\cellcolor[HTML]{FFFFFF}transistor sizing} & \multicolumn{1}{r|}{\cellcolor[HTML]{FFFFFF}450nm/180nm = 2.5} \\ \hline \vspace{-6pt}
 &  &  \\ \hline
\end{tabular}%
\end{table}

\subsection{Cell Test}

\begin{figure*}
\centering
\subfloat[]
{
	\includegraphics[scale=.47]{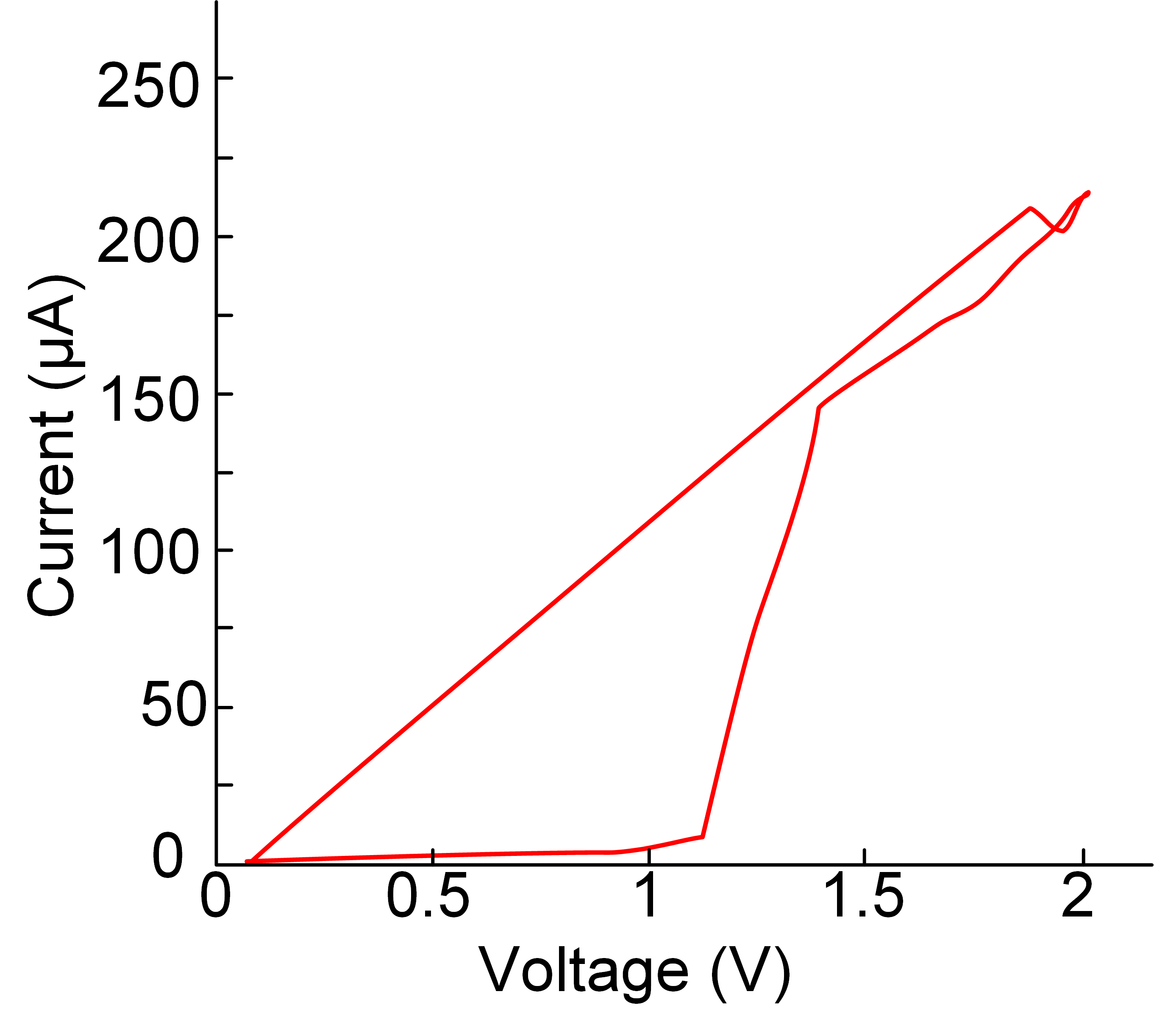}
	\label{fig:5a}
}
\subfloat[]
{
	\includegraphics[scale=.47]{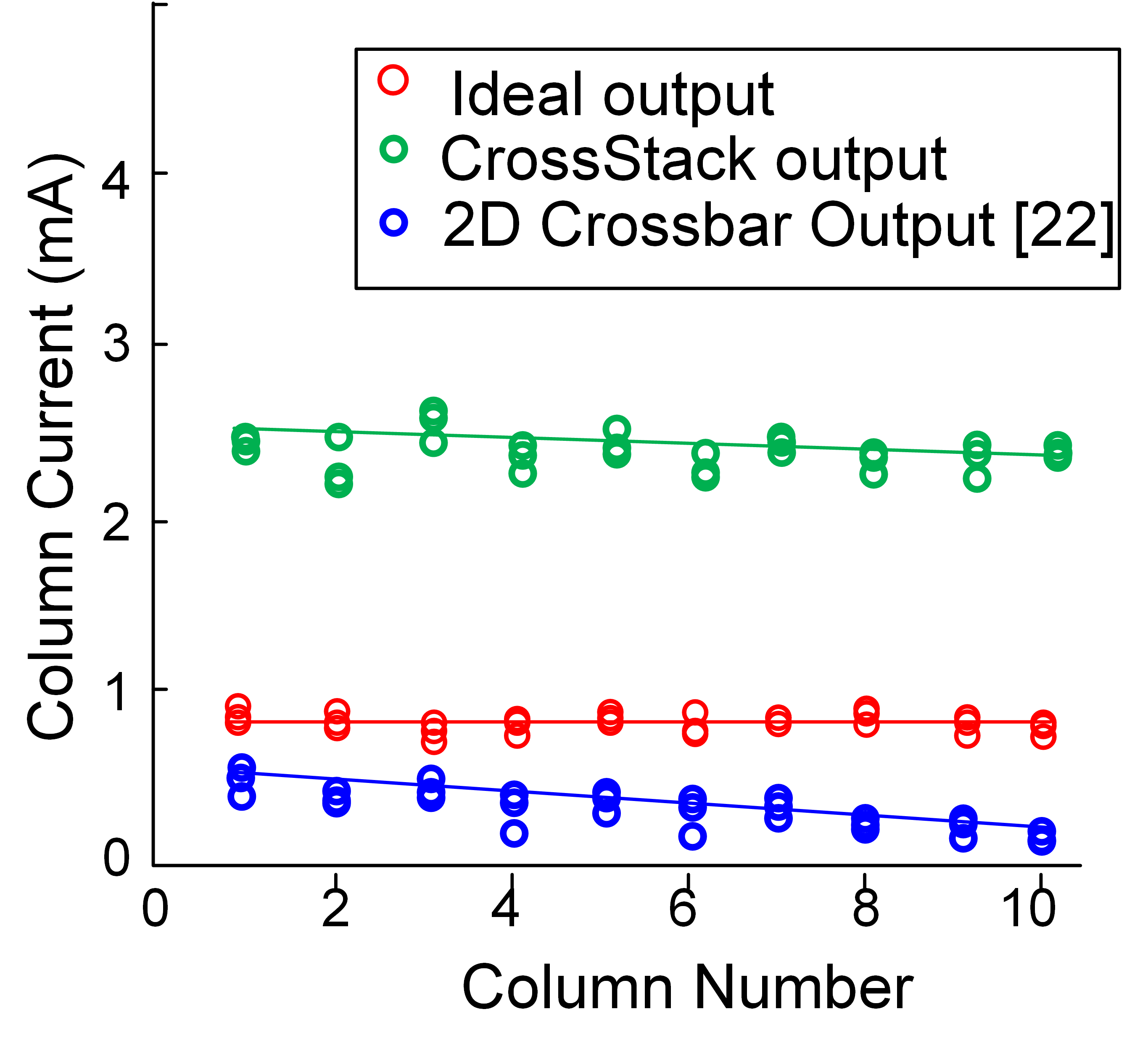}
	\label{fig:5b}
}
\subfloat[]
{
	\includegraphics[scale=.28]{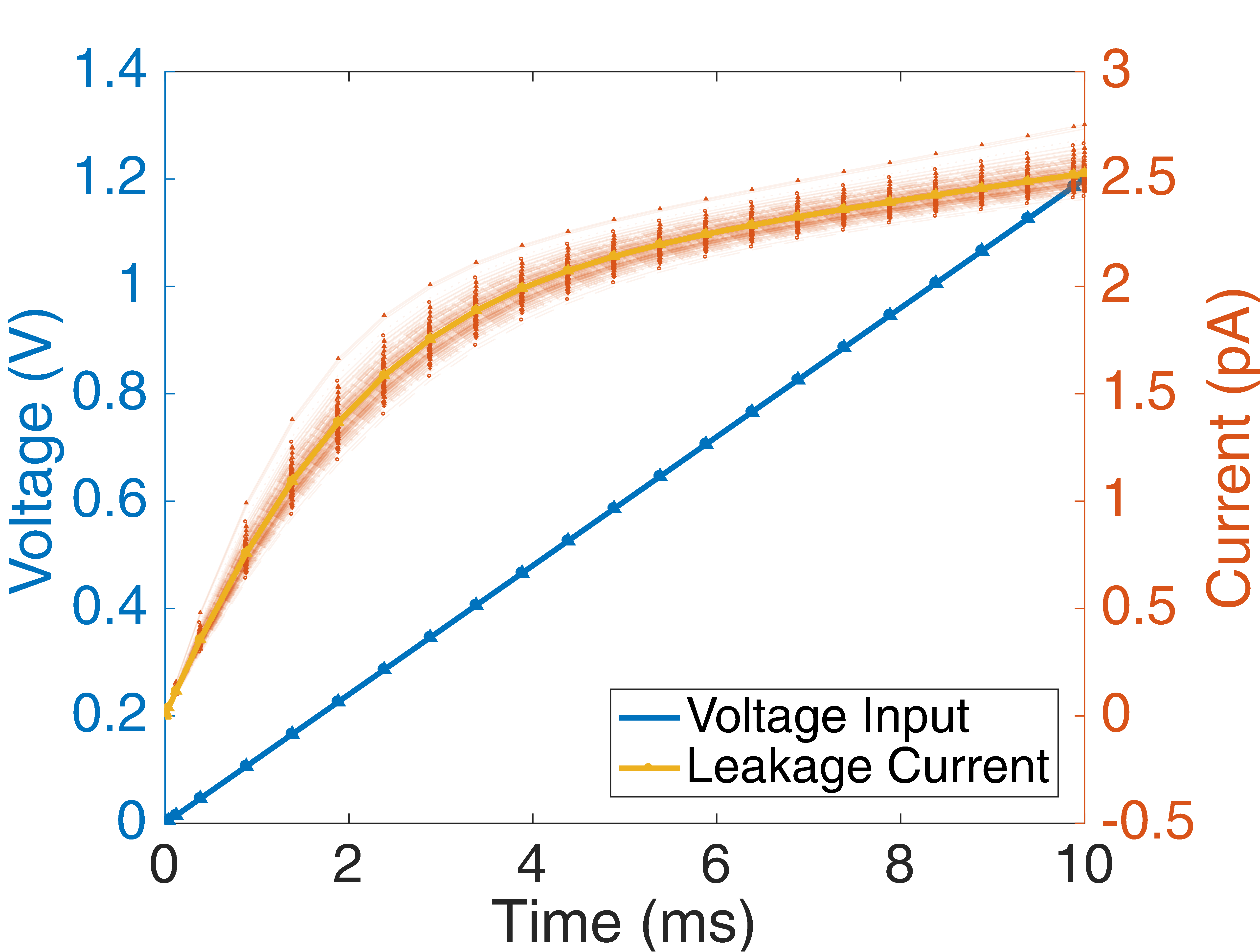}
	\label{fig:5a}
	}
	\subfloat[]
{
	\includegraphics[scale=.28]{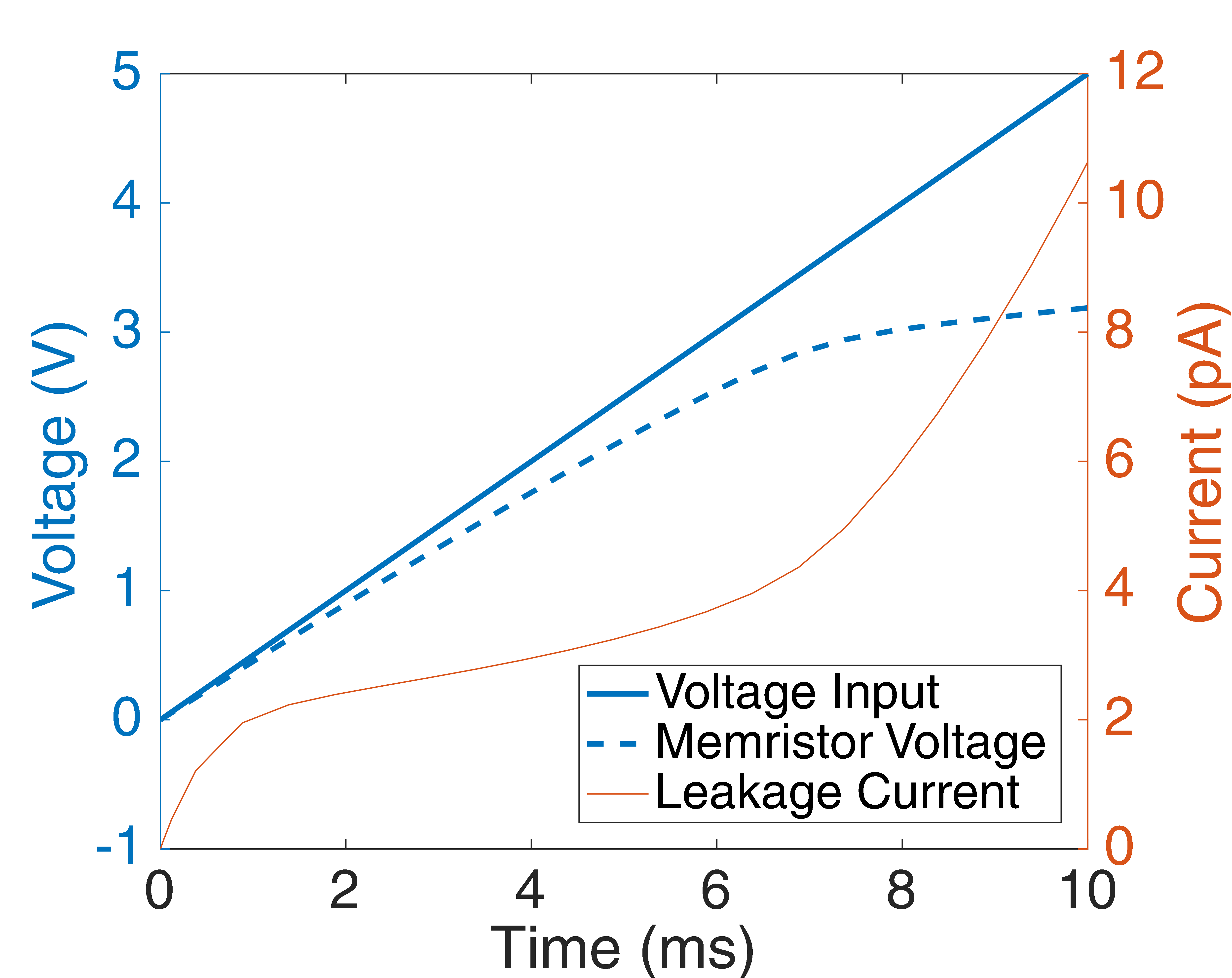}
	\label{fig:5a}
	}\\
\caption{Experimental results (a) Pinched hysteresis loop of a single memristor (b) IR loss comparison in expansion mode (c) Worst-case leakage current through transistor \texttt{N1} during a write cycle in deep-net mode with Monte Carlo parameter sweep of memristance (d) Extreme test case under large input write signal resulting in a nonlinear voltage drop across the memristor in deep-net mode.}
\label{fig:4}
\end{figure*}

The CMOS cell was designed in the SK Hynix 180-nm process where $v_{DD}$ = 1.8 V, $v_{Th}$ = 0.4 V, and our parameters are summarized in Table I. We used a read voltage in the range of \{0V, 0.5V\}, a write voltage of 1.2~V, both applied at $V_{IN}$, with measurements taken with a Micromanipulator tungsten probe tip. The pinched hysteresis loop measured with a 50Hz source is shown in Fig.~3(a). 

First, we tested the circuit in expansion mode. In the critical case of a write voltage applied to all devices $V_{write}$=1.2V and \texttt{RE} is set \texttt{HIGH} (1.8V), we show that IR drop is decreased by approximately 22\% compared to a similar planar inference engine in \cite{Hu2016}. This is shown by the slower decline of current output across columns in Fig.~3(b) for CrossStack, where the gold standard would be a perfectly straight line. Therefore, we verified that expansion mode reduces line losses for a given number of inputs due to the shorter length of column wire required. The trade-off is that column wires must handle twice the current capacity of an equivalent 2-D array. This demands wide column lines to handle such current capacity without risk of electromigration. But given that RRAM is integrated in the back end of the line, minimum thickness of higher layer routing wires may mitigate this.

Designing for deep-net mode opens up susceptibility to leakage currents through \texttt{N1} during write mode, concurrently with \texttt{N2} in read mode (see Fig.~1). The worst case leakage occurs along the shared column line, when there is a minimal read current and a maximal write current. The read array memristors will all be \texttt{OFF}, $R_r = 100K\Omega$, and all write array memristors will be \texttt{ON}, $R_s=10K\Omega$. To calculate the minimum value for $V_{read}$, we use 0.5~V as the maximum read voltage and assume an input of 7-bit resolution which requires increments of approximately $500mV/128 \approx 4mV$. The output current of a single cell under these conditions was measured to be 39.6~nA, which is 1\% off the ideal 40~nA. The accumulated leakage current through a column of 10 memristors being programmed ($V_{IN}$=1.2V; \texttt{RE=LOW}) was negligible in our experiments, and simulated to be approximately 2.5~pA per cell (i.e., 2.5pA$\times$10 cells = 25pA column current). This is $6.3\times10^{-2}$\% of the worst-case read-current, and so in the 180-nm process used, we are able to employ minimum transistor dimensions (W = 450nm, L = 180nm, W/L = 2.5). Leakage from a single cell is shown as a function of a DC sweep at the input is measured in Fig.~3(c) and (d), with a Monte Carlo parametric sweep of resistance overlaid (10k$\Omega\pm$7\%, Gaussian distribution across 200 trials).

\begin{figure}
\centering
\includegraphics[scale=0.6]{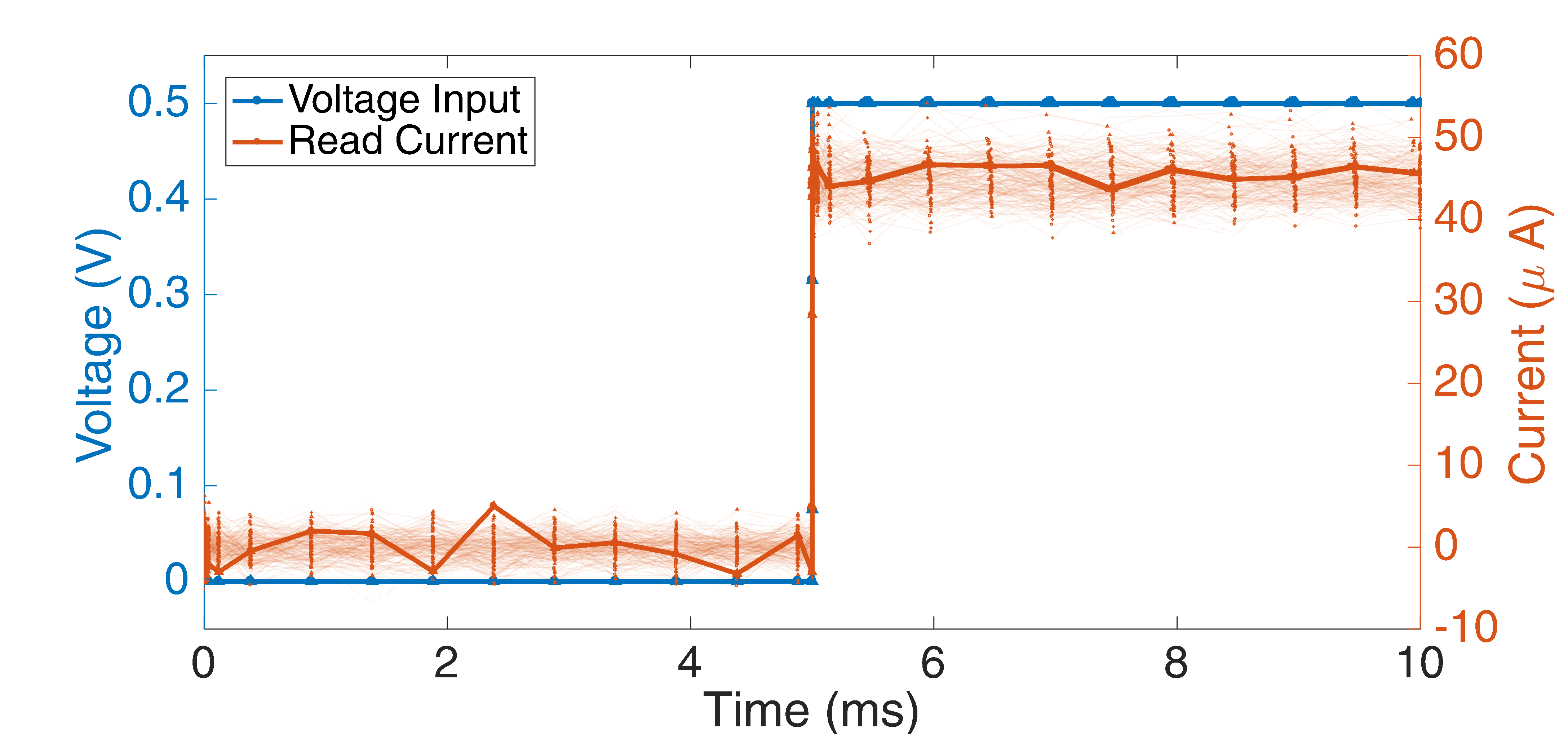}
\caption{Transient analysis of a single cell current output during deep-net mode in a read cycle.}
\end{figure}

Current read-out measurements taken from a transient analysis under a switching input during a read-cycle are given in Fig.~4, where current deviates by 8\% from the measured value in the worst-case. This suggests that a single device should be limited to 3.5-bits. We note this is a limitation not of the CrossStack architecture, but rather of memristor variability. In the conservative case of a 1-bit cell across 10 columns, the ~10ns read-out is subsumed within the 25ns programming time as opposed to being added to a separate cycle after programming. This confirms improvement in speed due to our in-situ pipelining mechanism by 29\%. 




\section{Discussion and Conclusion}
With two different modes available to the user, how would you decide to use one over another? In short, deep-net is suited for tasks where speed is paramount and expansion mode is for tasks involving a large number of inputs, be it for increased resolution, or to process a very large number of parameters in a shallow network.

There are three primary motivations in the use of expansion mode: 1) fully-connected networks typically have a far larger number of connections than in convolutional layers. Expansion mode doubles the number of possible inputs for a fixed area which enables increases the number of possible inputs; 2) a larger number of inputs requires a larger length of column wire. By using expansion mode to double up on inputs, we reduce wire resistance to half the original amount for a given number of inputs. More inputs per unit length of wire enables our crossbar to be resilient to write failures arising from line resistance IR drops by reducing line losses by 22\%. 3) 
The lack of reliable analog memory technology makes it hard to perform hardware multiplexing in analog, and transmitting analog values over long distances or at high speed is not efficient. Restricting each memristor to one of two conductance values (i.e., single-bit memristors) means that one would require $log_2(n)$ memristors for $n$ bits of precision. For crossbars that use conductance states of memristors conservatively, digital computing is more desirable but requires more memristors in a crossbar than analog for the same precision. Expansion mode facilitates this increase in devices whilst halving crossbar area. The drawback is that a larger current must be carried through the wire with more vias.

Deep-net mode is a novel processing scheme where the two crossbar layers are isolated from one another by appropriately biasing \texttt{RE}, in order to parallelize read and write operations. It is engaged where speed is of greater importance than precision. In the most simplistic way (ignoring max-pooling and dropout), a crossbar performs inference in the following way:
\begin{enumerate}
\item Write weights to memristor conductances,
\item Apply a sub-threshold read voltage,
\item Buffer or store read-out signal in memory,
\item Write the next hidden layer weights to the crossbar,
\item Repeat steps 2-5 until output is generated.
\end{enumerate}

\noindent In deep-net mode, CrossStack performs steps 2 and 4 simultaneously which enables read and write operations to occur together. By the time an output is generated from step 3, it is ready to be processed by the next hidden layer in step 4 for a speed increase of 29\% over an equivalent 2-D array. 

The most prevailing issues in realizing large scale crossbar arrays beyond our working prototype are mismatch and endurance. At this stage, our current error rate was 8\% which limits the number of bits that can be represented by a single cell. Subthreshold current was hardly an issue in our design, but Fig.~3(d) shows the nonlinearity of voltage across the memristor under high write voltages ($V_{IN}>3V$), and leakages will become more prevalent at shorter channel lengths. Capacitive coupling as a result of high programming voltages into bit-lines that are reading out poses a degree of risk, but should be tolerable for higher metal layers given the wider spacing. Allowing for heat dissipation in a 3D structure is an ongoing challenge and the subject of significant process-related research, often calling for external heat sinks. In general, the advantages seem to outweigh the drawbacks and CrossStack presents a promising methodology for reconfigurable inference acceleration to adapt to the various types of ANNs being deployed.

\section*{Acknowledgment}
This work was supported by the National Research Foundation of Korea grant funded by the Korean government (MSIT) (No.2020R1F1A1069381).





%

\end{document}